\documentclass[%
 reprint,showpacs,
superscriptaddress,
 amsmath,amssymb,
 aps,
prb,
floatfix
]{revtex4-1}
\usepackage{amssymb, blindtext, amsmath, lineno, placeins, graphicx, capt-of}

\usepackage{graphicx}
\usepackage{dcolumn}
\usepackage{bm}
\usepackage{float}

\usepackage{lineno}
\newcommand{\mv}[1]{\ensuremath{{\mathbf #1}}}

\newcommand{\curl}{\nabla\times}

\newcommand{\dslab}{d_{\text{slab}}}
\newcommand{\tstar}{{t^{\ast}}}
\newcommand{\ee}{\mathrm{e}}
\newcommand{\ii}{\mathrm{i}}
\newcommand{\pol}{{(\text{P})}}
\newcommand{\poln}{{(-1)^{\scriptsize{\text{P}}}}}
\newcommand{\polnm}{{(-1)^{\scriptsize{\text{P}}}}}

\begin{document}


\title{Retrieving effective material parameters of metamaterials characterized by nonlocal constitutive relations}

\author{Karim Mnasri}%
 \affiliation{Institute of Theoretical Solid State Physics, Karlsruhe Institute of Technology, Wolfgang-Gaede-Str. 1 76131 Karlsruhe Germany}%
 \email{karim.mnasri@kit.edu}

\author{Andrii Khrabustovskyi}%
\affiliation{Institute of Applied Mathematics, Graz University of Technology, Steyrergasse 30 8010 Graz Austria}%

\author{Michael Plum}%
\affiliation{Institute for Analysis, Karlsruhe Institute of Technology, Englerstr. 2 76131 Karlsruhe Germany}%

\author{Carsten Rockstuhl}%
 \affiliation{Institute of Theoretical Solid State Physics, Karlsruhe Institute of Technology, Wolfgang-Gaede-Str. 1 76131 Karlsruhe Germany}%
\affiliation{Institute of Nanotechnology, Karlsruhe Institute of Technology, P.O. Box 3640 76021 Karlsruhe Germany}%

\date{\today}

\begin{abstract}
The parameter retrieval is a procedure in which effective material properties are assigned to a given metamaterial. A widely used technique bases on the inversion of reflection and transmission from a metamaterial slab. Thus far, local constitutive relations have been frequently considered in this retrieval procedure to describe the metamaterial at the effective level. This, however, is insufficient. The retrieved local material properties frequently fail to predict reliably the optical response from the slab in situations that deviate from those that have been considered in the retrieval, e.g. when illuminating the slab at a different incidence angle. To significantly improve the situation, we describe here a parameter retrieval, also based on the inversion of reflection and transmission from a slab, that describes the metamaterial at the effective level with nonlocal constitutive relations. We retrieve the effective material parameters at the example of a basic metamaterial, namely dielectric spheres on a cubic lattice but also on a more advanced, anisotropic metamaterial of current interest, i.e., the fishnet metamaterial. We demonstrate that the nonlocal constitutive relation can describe the optical response much better than local constitutive relation would do. Our approach is widely applicable to a large class of metamaterials.  
\begin{description}
\item[PACS numbers] 41.20.Jb,42.70.-a,78.20.Bh,78.20.Ci,78.67.Pt
\end{description}
\end{abstract}

\maketitle

\section{\label{sec:introduction}Introduction}
Optical metamaterials constitute a novel class of materials that can control the propagation of light in a way inaccessible with natural materials \cite{APPLICATION1, APPLICATION2, APPLICATION3, APPLICATION4, APPLICATION5}. Optical metamaterials are mostly complicated structures with a spatially distributed permittivity $\epsilon(x,y,z,k_0)$ that is, very often, periodic in space. To describe the propagation of light through such materials, full-wave numerical solvers of Maxwell's equations that take into account all the fine details of the spatially dependent permittivity are always an option. Examples for such numerical solvers are the Fourier Modal Method (FMM) \cite{FMMREF,FMMREF2}, the finite element method (FEM) \cite{FEMREF}, or the finite-difference time-domain (FDTD) method \cite{FDTDREF}. These approaches share the heavy request on computational resources. Therefore, to effectively consider optical metamaterials in the design of functional devices, we shall not describe them at the mesoscopic level, i.e., while considering the fine details of the unit cell, but rather at an effective level. By treating them as effectively homogeneous, we put metamaterials on an equal footing to ordinary materials. To make this homogenization, the assignment of effective material parameters is of paramount importance. This is done in a process called the parameter retrieval. 
 
The starting step in the parameter retrieval is the agreement on a particular constitutive relation that shall describe the metamaterial at the effective level. Frequently, for centro-symmetric material with no magneto-electric coupling on which we will concentrate here and inspired by the way we treat natural materials, local constitutive relations are assumed \cite{simovski2010electromagnetic,PhysRevLett.98.037403}, i.e., ${\mv D (\mv r, k_0) =\epsilon (k_0) \mv E(\mv r, k_0)}$ and ${\mv B(\mv r, k_0) = \mu(k_0) \mv H(\mv r, k_0)}$. The effective material parameters are the electric permittivity $\epsilon(k_0)$ and the magnetic permeability $\mu(k_0)$. $k_0=\frac{\omega}{c}$ is the free space wavenumber and $\omega$ and $c$ are the frequency of the considered time-harmonic field and the speed of light in vacuum, respectively. We refer to the description of a metamaterial with these two parameters only, as the Weak Spatial Dispersion (WSD) or local approximation. It is a local constitutive relation since the electric displacement $\mv D (\mv r, k_0)$ and the magnetic induction $\mv B (\mv r, k_0)$ depend only locally on the electric field $\mv E (\mv r, k_0)$ and the magnetic field $\mv H (\mv r, k_0)$, respectively.

However, metamaterials are usually made from building blocks, also called meta-atoms, that have a size in the order of several tens or even hundreds of nanometers, while being designed to operate at optical or near-infrared wavelengths. This is in stark contrast to natural materials that have critical dimensions of merely a fraction of one nanometer. The disparate length scales between critical feature size and operational wavelength for natural materials justifies their treatment with local constitutive relations. Indeed, it is quite a challenge to trace signatures of a nonlocal character with natural materials \cite{PhysRevB.91.184207,AGRANOVICH20091}. The assumption of a local medium, however, ceases to be applicable for optical metamaterials when their critical length scale is no longer much smaller than the wavelength but only smaller. Then, nonlocal effects can no longer be neglected. 

But how can we judge, which effective description is appropriate? Well, first of all and with the purpose to treat the material as effectively homogeneous, we require it to be sub-wavelength under all circumstances. A frequent situation, adapted in its geometry also to experimental constraints, is the availability of the metamaterial as a slab with a finite thickness. To perceive such metamaterial from the outer world as homogeneous, we require the abscence of a first diffraction order at oblique incidences. This requires the period $a$ of the metamaterial to be smaller than half the operational wavelength $\lambda$, i.e., $ \lambda > 2a$ (cf. Ref.~\onlinecite{PhilippeLalanne}).

Next, in the retrieval procedure the reflection and transmission from the metamaterial illuminated with a linearly polarized plane wave are calculated with a full wave solver. By inverting the expressions for reflection and transmission from a homogeneous slab characterized by a specific constitutive relation, the effective material parameters can be retrieved. However, these retrieved properties shall predict the optical response from the same metamaterial also in situations that have not been considered in the retrieval. If they fail, the effective material properties would be useful to reproduce reflection and transmission for the same situation in which they have been retrieved but they could not be used for anything else. This would be strongly against the idea of a material parameter. 
     
When considering local constitutive relation, it has been shown in the past, and for comparative purpose we also show below at a specific example for a metamaterial, that the retrieved local material parameters are insufficient to cope with this requirement \cite{PhysRevLett.115.177402,0957-4484-26-18-184001}. Local constitutive relations can reasonably explain the optical response for waves close to near-normal incidence but they fail to describe the optical response at angles beyond the paraxial regime \cite{PhysRevB.81.035320}. The situation is of course more severe at wavelengths close to the resonance. These are clear indications that local constitutive relations are insufficient to capture the optical response from metamaterials. Instead, nonlocal constitutive relations have to be considered. The general importance of considering nonlocality for a reasonable parameter retrieval has been also pointed out by other authors and can be considered to be accepted by now in the literature \cite{Tsukerman:11,PhysRevB.84.075153,photonics2020365,PhysRevB.86.085146,PhysRevB.92.085107,PhysRevB.67.113103,PhysRevB.92.045127}. The possibilities of deriving (additional) interface conditions for the macroscopic fields are numerous. A rigorous approach for deriving additional interface conditions that are, however, only applicable for wire media structures may be found in Refs.\onlinecite{PhysRevE.73.04661} and \onlinecite{1638373}. Using first principle approach, the author in Ref.~\onlinecite{1367-2630-16-8-083042} derives additional interface conditions for an arbitrary material with a quatdrupolar-type response.

In the past, we introduced two closed forms of nonlocal constitutive relations and showed that they can capture very well the bulk properties of a given fishnet metamaterial \cite{PREVIOUSPAPER}. In particular, the dispersion relation of the eigenmodes of the fishnet metamaterial could be correctly reproduced with the nonlocal constitutive relations. We investigated, on the one hand, a second-order model for the nonlocality that retains all symmetry allowed terms of the fishnet structure ($\mathrm{D}_{2\mathrm{h}}$-symmetry) and, on the other hand, a specific fourth-order model for the nonlocality. Since the fourth-order model offered a better description of the bulk properties, we decided to investigate this model in this manuscript as well. We emphasize that, besides these two models, there are many more models that could have been considered, in general. Selecting one or another model for the constitutive relation is a sensitive task. However, we have based our decision to study a specific model on (a) the requirement to introduce a rather small number of additional parameters at the level of the effective description to avoid too much arbitrariness, (b) the ability of the additionally considered terms to unlock typical features for strong spatial dispersion, (c) the feasibility to study the dispersion relation and the possibility to derive interface condition with a specific constitutive relation, and (d) the ability to link a specifically chosen functional dependency for the constitutive relation to those that have been previously considered. All these aspects are met by the aforementioned fourth-order model.  

To significantly advance this approach, we outline here a procedure to retrieve the actual nonlocal material parameters. We rely for this purpose on reproducing the reflection and transmission coefficients from a slab of a given metamaterial with those calculated under the assumption of a homogeneous but nonlocal metamaterial. We show that the retrieved nonlocal material parameter can correctly describe the metamaterial at the effective level beyond the paraxial regime. We also show that artefacts that have been controversially discussed in the initial research period on metamaterials, e.g. anti-Lorentz resonances and a negative imaginary part in the effective permittivity\cite{PhysRevE.68.065602,PhysRevE.71.036617} at the wavelength of the magnetic resonance, vanish when nonlocal constitutive relations are considered. Therefore, as often speculated but now demonstrated, we deem these artefacts to be associated to the non-adapted description of the metamaterial at the effective level with local constitutive relations. We stress upfront that while we can mitigate these features from the local material parameters, we continue to encounter them in the nonlocal material parameters. At the moment, we can only speculate that when considering higher order terms in the nonlocal terms, these problems in lower order terms will vanish as well.      

In the following Sec.~\ref{sec:setup} we discuss the basic system considered and describe all the theoretical background to predict the optical response from a slab of a homogeneous metamaterial characterized by non-local constitutive relations. The technical details of the actual retrieval procedure will be outlined in Sec.~\ref{sec:retrieval-procedure}. We retrieve angle-independent material parameters for a fishnet metamaterial also in Sec.~\ref{sec:retrieval-procedure} and discuss them in depth. Prior to the fishnet, we discuss an easier example of a metamaterial made by isotropic unit cells, namely dielectric spheres on a cubic lattice. We compare our results to those obtained with a local model and quantify the improvement. Finally, we conclude and summarize our work in Sec.~\ref{sec:conclusions}.

\newpage
\begin{widetext}

\section{\label{sec:setup}Setting up the Fresnel equations}
\begin{figure}[h]
	\centering
  \includegraphics[width=0.4\textwidth]{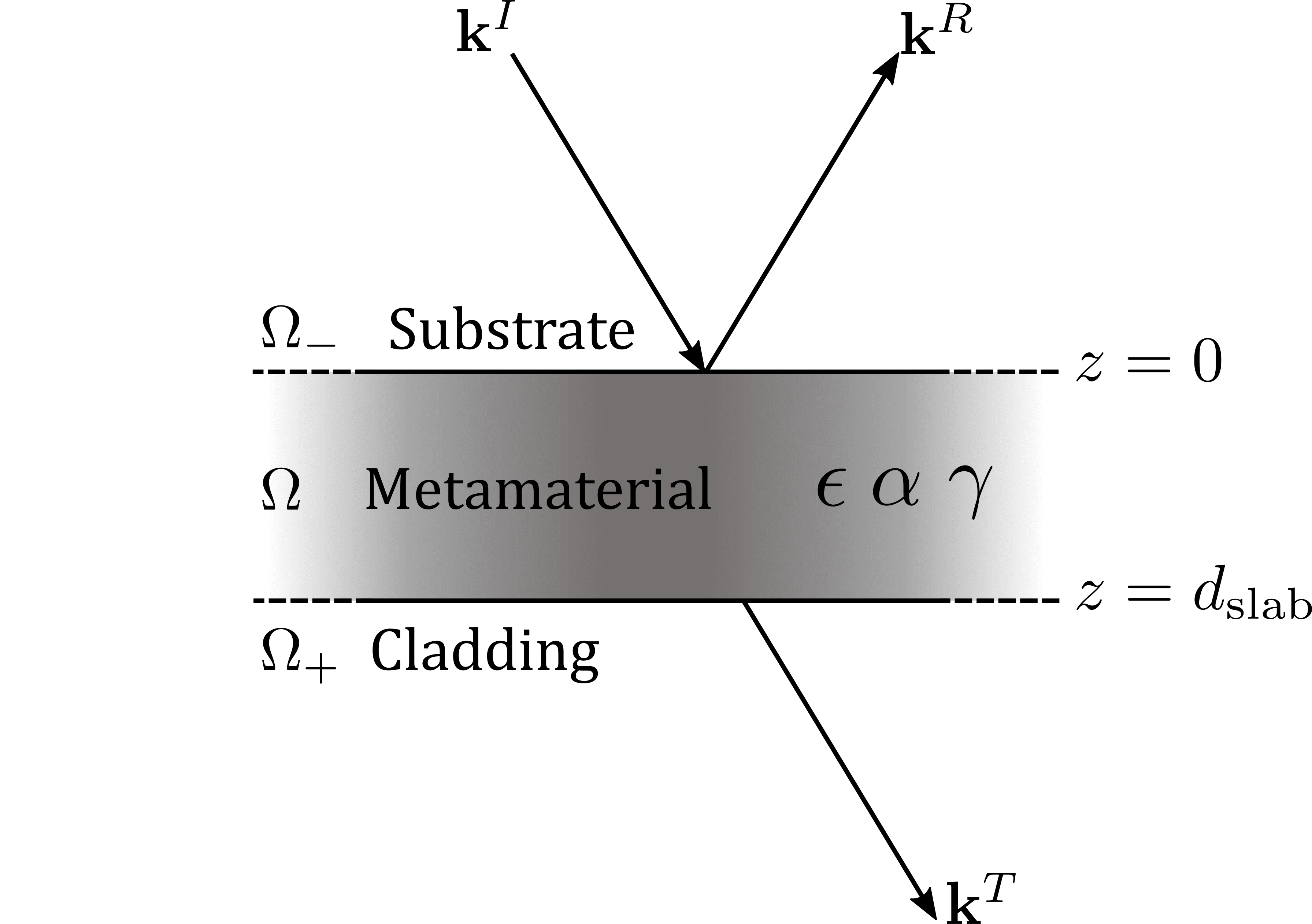}
	\caption{Basic geometry of the slab problem. The metamaterial slab is infinitely extended in the $xy$-plane with a thickness $d_{\text{slab}}$ in $z$-direction, defining the region $\Omega$. The incident and transmission half-spaces $\Omega_-$ and $\Omega_+$, respectively, are assumed to be air.}
	\label{fig:basic slab}
\end{figure} 
For the purpose of homogenizing centro-symmetric metamaterials with nonlocal constitutive relations, let us consider a nontrivial, but homogeneous material being infinitely extended in the $xy$-plane with a thickness $d_{\text{slab}}$ in $z$-direction. It shall fill the space $\Omega:=\mathbb{R}^2 \times(0, \dslab)$. The incident light $(\mv k^I, \mv E^I,\mv H^I)$ and the reflected light $(\mv k^R, \mv E^R,\mv H^R)$ reside in the incidence half-space $\Omega_- := \mathbb{R}^2\times(-\infty,0)$ and the transmitted light $(\mv k^T, \mv E^T,\mv H^T)$ in the transmission half-space $\Omega_+:=\mathbb{R}^2\times(\dslab,\infty)$. The half-spaces $\Omega_-$ and $\Omega_+$ shall later represent the substrate and cladding, respectively, and be both filled with air ($\epsilon=\mu=1$). Of course, other materials could be considered as well. With the objective to homogenization, a displacement field $\mv D$, in the presence of light with an electric field  $\mv E$, has been lately introduced in Ref.~\onlinecite{PREVIOUSPAPER} in the frequency domain that reads
\begin{align} \label{eq:const-relations}
    \mv D[\mv E] = \begin{cases}
                    \mv E \qquad \text{for } \mv r \in \Omega_- \cup \Omega_+ \,,\\ 
                    \epsilon   \mv E + \curl \alpha \left(\curl\mv E\right)
                    + \curl\curl \gamma \left(\curl\curl\mv E\right)  \qquad \text{for } \mv r \in \Omega \,,
    \end{cases}
\end{align}
with the tensorial and frequency-dependent effective material parameters $\epsilon$, $\alpha$, and $\gamma$. We skip here and in the following the frequency and space dependency in the arguments to simplify the notation. However, it is implicitly considered. We would like to emphasize that in region $\Omega$ the effective response tensor in Fourier-space $\hat{\mv{R}}(\mv{k})$, with $ \mv D[\mv E] = \hat{\mv{R}}(\mv{k})\cdot \mv E $, also contains off-diagonal tensor elements, even in an isotropic medium. In this case the effective material parameters $\epsilon$, $\alpha$, and $\gamma$ become scalars, but $\hat{\mv{R}}(\mv{k})$ will remain dense with off-diagonal terms scaling as $\mv k\mv k$ (cf. Eq.(7) in Ref.~\onlinecite{PhysRevB.92.045127}).
Here, we assume that the material is intrinsically nonmagnetic, such that $\mv{B}(\mv r, k_0)=\mv{H}(\mv r, k_0)$. However, as a consequence of the finite size and the sophisticated geometry of metamtaterial's building-blocks, currents in a closed loop can be induced. They lead to an artificial effective magnetic response that is linked to the second term in the constitutive relation of the electric field above via $\alpha_i = \frac{\mu_i-1}{k_0^2\mu_i}$, where $i$ refers to a spatial coordinate. Hence, the parameter $\alpha$ can be re-interpreted as a local, effective magnetic permeability $\mu$ and potentially leads to a negative index behavior. As this second term can be recast to appear as a local magnetic response, the model with $\epsilon$ and $\alpha$ only, or alternatively $\epsilon$ and $\mu$ only, will be denoted as the Weak Spatial Dispersion (WSD) or the local model. In contrast, the model that also includes the nonlocal material parameter $\gamma$ will be denoted as the Strong Spatial Dispersion (SSD) or the nonlocal model. Throughout this manuscript, we will treat both WSD and SSD models simultaneously, with careful consideration of the limit $\gamma \rightarrow 0 $.

For convenience, we align the laboratory frame with the principal axes of the metamaterial. The material parameters are, therefore, diagonal and read
\begin{align}\label{eq:coeff}
\begin{array}{cccccc}
&\epsilon =\left(\begin{matrix}
\epsilon_x&0&0\\0&\epsilon_y&0\\0&0&\epsilon_z
\end{matrix}\right),
&\alpha =\left(\begin{matrix}
\alpha_x&0&0\\0&\alpha_y&0\\0&0&\alpha_z
\end{matrix}\right),  
&\gamma =\left(\begin{matrix}
\gamma_x&0&0\\0&\gamma_y&0\\0&0&\gamma_z
\end{matrix}\right)\,.
\end{array}
\end{align}
For technical convenience and without loss of generality, we shall assume that the incident plane is either in the $xz$-plane or in the $yz$-plane and the propagation direction is normal to the slab, i.e., in the positive $z$-direction. The incident wave-vector is, therefore, either $\mv k^I=(k_x,0,k_z)$ or $\mv k^I=(0,k_y,k_z)$. In addition, since the material is centro-symmetric, no optical activity takes place and the polarization of the fields is preserved. It is thus sufficient to consider linearly polarized waves only and to decompose the eigenmodes into decoupled TE (transverse electric) and TM (transverse magnetic) modes. For the sake of notation, we shall denote the transverse component of $\mv k$ by $k_t$, where $t$ is either $x$ or $y$ and $\tstar=x$ for $t=y$ and and $\tstar=y$ for $t=x$. 

In the regions outside the slab, i.e., in $\Omega_- \cup \Omega_+$ the dispersion relation is $\left(k_z^{(i)}(k_0,k_t)\right)^2= k_0^2-k_t^2$, where the superscript $i\in\{I,R,T\}$ represents the incident, reflected, or transmitted fields, respectively. In region $\Omega$, the solutions to the wave equation, i.e., the dispersion relations $\left(k_{z,\sigma}(k_0,k_t)\right)^2$ are multiple, with $\sigma=\pm 1$, are derived and discussed in depth in Ref.~\onlinecite{PREVIOUSPAPER}. The polarizations in this case differ remarkably such that the TE and TM polarizations show different functional dependency of $k_{z,\sigma}(k_0,k_t)$. This also severely affects the Fresnel equations and finally the reflection and transmission coefficients. In the case of the TE polarization the dispersion relation reads
\begin{align}
     \left(k_{z,\sigma}^{\text{TE}}(k_0,k_t)\right)^2  =& -k_t^2+p_0^{\text{TE}} + \sigma \sqrt{\left(p_0^{\text{TE}}\right)^2-q_1^{\text{TE}}+2\left(p_1^{\text{TE}}-p_0^{\text{TE}}\right)k_t^2} \label{eq:dispSSDTE}
\end{align}
with $\sigma=\pm 1$ and the coefficients $p_0^{\text{TE}} = [2k_0^2\gamma_{\tstar}\mu_t]^{-1} \,$,    $p_1^{\text{TE}} = [2k_0^2\gamma_{\tstar}\mu_z]^{-1}\,$ and  $q_1^{\text{TE}} = \frac{\epsilon_{\tstar}}{\gamma_{\tstar}}\,$.
Whereas for the TM polarized field we obtain
\begin{align}
   \left(k_{z,\sigma}^{\text{TM}}(k_0,k_t)\right)^2  =& -\frac{1}{2}\left( q_0^{\text{TM}}+q_1^{\text{TM}}\right) k_t^2 +p_0^{\text{TM}} + \sigma
                \sqrt{\left(p_0^{\text{TM}}+\frac{q_0^{\text{TM}}-q_1^{\text{TM}}}{2}k_t^2\right)^2-p_1^{\text{TM}}} \label{eq:dispSSDTM}
\end{align}
with $\sigma=\pm 1$ and the coefficients
     $p_0^{\text{TM}} = [2k_0^2\mu_{\tstar}\gamma_t]^{-1}\,$,
     $p_1^{\text{TM}} = \frac{\epsilon_t}{\gamma_t}\,$.
     $q_0^{\text{TM}} = \frac{\epsilon_t}{\epsilon_z} \,$ and 
     $q_1^{\text{TM}} = \frac{\gamma_z}{\gamma_t}\,$.
We shortly want to recall that the dispersion relation for a local medium can be reproduced by considering the limit $\gamma\rightarrow 0 $. Here, one has to be just careful since one of the $k_{z,\sigma}$ asymptotically behaves like $\frac{1}{\sqrt{\gamma}}$ as $\gamma\rightarrow 0$. Without loss of generality let $k_{z,-}$ be the divergent solution. 

With the information above, one can solve the bulk problem and retrieve the wave parameters, i.e., dispersion relation and the effective refractive index. Despite that, the real utility of the effective medium description of a metamaterial is not to reproduce the wave parameters only, but rather, the real utility of effective medium description is also in relating the structure of a metamaterial to its transmission and reflection and to retrieve the effective material parameters. It is therefore important to understand reflection and transmission of light through a slab of a metamaterial surrounded by air. To this end, and after we have set up the pieces above together, we can write down the Fresnel matrix that has been rigorously derived for the model \eqref{eq:const-relations} in Ref.~\onlinecite{PREVIOUSPAPER}
\begin{equation}
     \mv {F}^\pol = \left(
\begin{array}{cccccc}
 r_1^\pol & a_+^\pol & a_-^\pol &  a_+^\pol \poln&  a_-^\pol \poln& 0 \\
 r_2^\pol & b_+^\pol & b_-^\pol & -b_+^\pol \polnm  & -b_-^\pol \polnm  & 0 \\
 0        & c_+^\pol & c_-^\pol &c_+^\pol  \poln &  c_-^\pol \poln& 0 \\
0  & a_+^\pol\ee^{\ii \psi_+^\pol} & a_-^\pol\ee^{\ii \psi_-^\pol} &  a_+^\pol\poln\ee^{-\ii \psi_+^\pol} & a_-^\pol\poln\ee^{-\ii \psi_-^\pol} & t_1^\pol \\
0  & b_+^\pol\ee^{\ii \psi_+^\pol} & b_-^\pol\ee^{\ii \psi_-^\pol} & -b_+^\pol \polnm\ee^{-\ii \psi_+^\pol}  & -b_-^\pol\polnm \ee^{-\ii \psi_-^\pol}  & t_2^\pol \\
 0        & c_+^\pol\ee^{\ii \psi_+^\pol} & c_-^\pol\ee^{\ii \psi_-^\pol} &c_+^\pol \poln\ee^{-\ii \psi_+^\pol}  &  c_-^\pol\poln\ee^{-\ii \psi_-^\pol} & 0 
\end{array} \right)
\end{equation}
and $\mv I^\pol = (i_{1}^\pol,i_{2}^\pol,0,0,0,0)^{\text{T}}$. The matrix elements are very long, and therefore, summarized in Tab.~\ref{table:fresnelcoeffsTETM}. We note that P indicates the polarization. The Fresnel matrix above has to be understood as follows. The first three rows correspond to the interface conditions at the first interface at $z=0$ and the last three ones to the second interface of the slab, where $z=\dslab$ and the fields accumulate a phase $\ee^{\ii \psi_\sigma^\pol}$, where $\psi_\sigma^\pol = k_{z,\sigma}^\pol\dslab$. The first and sixth columns contain information regarding the reflected and transmitted fields, respectively. The second and fourth columns pertain the terms for the forward propagating modes inside a slab with $\Im (k_{z,\sigma}^\pol)>0$ while the third and fifth columns are associated to the modes that propagate backwards inside the slab with $\Im (k_{z,\sigma}^\pol)<0$. To reconstruct the Fresnel matrix for the WSD, one has to consider the limit $\gamma \rightarrow 0$. Two rows and columns will contain the divergent $k_{z,-}^\pol$, which exponentially damps the field amplitudes and, therefore, do not contribute neither to reflection nor to transmission. The effective dimension of the Fresnel matrix in the case of WSD reduces to a $4\times 4$-matrix, as expected. 

Using the Fresnel matrix above, we obtain the complex-valued reflection and transmission coefficients by taking the first and the last components, respectively. Hence,
\begin{align}
    \rho^\pol(k_0,k_t,\epsilon,\mu,\gamma) &= \left[\left(\mv F^\pol\right)^{-1} \cdot \mv I^\pol\right]_1 \,, \label{eq:reflection-definition} \\
    \tau^\pol(k_0,k_t,\epsilon,\mu,\gamma) &=  \left[\left(\mv F^\pol\right)^{-1} \cdot \mv I^\pol\right]_6 \,. \label{eq:transmission-definition} 
\end{align}
The formulas for both $\rho^\pol$ and $\tau^\pol$ are very long and will, for the sake, of readability not written explicitly. Yet, they will be used in the next section for the parameter retrieval and evaluated for comparison to the reflection and transmission coefficient of the actual material to be homogenized.

\begin{center}
\captionof{table}{Substitution table for the Fresnel coefficients for both TE and TM polarization and fixed incident plane $tz$.} \label{table:fresnelcoeffsTETM}
 \begin{tabular}{|c| c| c|} 
 \hline
   & $P=0$ (TE) & $P=1$ (TM) \\ [0.5ex] 
 \hline\hline
 $a_{\sigma}^\pol$ & $-1$ & $-\frac{\epsilon_z}{\epsilon_t}k_{z,\sigma}$ \\ 
 \hline
 $b_{\sigma}^\pol$ & $k_{z,\sigma}\left[k_0^2 \gamma_\tstar\left( k_t^2+k_{z,\sigma}^2\right)-\mu_t^{-1} \right]$  & $\left[ k_0^2\left(\gamma_z k_t^2+\gamma_t k_{z,\sigma}^2\right) -\mu_\tstar^{-1}
 \right]\left(k_t^2+ \frac{\epsilon_z}{\epsilon_t}k_{z,\sigma}^2 \right)$  \\
 \hline
 $c_{\sigma}^\pol$ & $\gamma_\tstar\left(k_t^2+k_{z,\sigma}^2\right)$ & $\gamma_t k_{z,\sigma}\left( k_t^2+ \frac{\epsilon_z}{\epsilon_t}k_{z,\sigma}^2 \right)$   \\
 \hline
 $\psi_{\sigma}^\pol$ & $k_{z,\sigma} \dslab$ & $k_{z,\sigma} \dslab$ \\ 
 \hline
 $r_{1}^\pol$ & $1$ & $k_z^R$ \\
  \hline
 $r_{2}^\pol$ & $k_z^R$ & $k_0^2$ \\
 \hline
 $t_{1}^\pol$ & $1$ &  $k_z^T$ \\
 \hline
 $t_{2}^\pol$ & $k_z^T$ &  $k_0^2$ \\
 \hline
 $i_{1}^\pol$ & $-1$ & $-k_z^I$ \\
  \hline
 $i_{2}^\pol$ & $-k_z^I$ & $-k_0^2$ \\[1ex] 
 \hline
\end{tabular}
\end{center}
\end{widetext}

\newpage
\section{\label{sec:retrieval-procedure}Retrieval procedure}

In general, light-matter interaction depends on the polarization and the incident plane of the light. Therefore, the electromagnetic fields couple to different material parameters in each scenario. Due to the anisotropy of the structure, one can only retrieve all effective material parameters by considering all four possible illuminations (TE,TM)$\times(k_x,k_y)$. The relevant material parameters for a fixed illumination condition are summarized in Tab.~\ref{table:materialparamters}. The nonlocal parameter $\gamma$ behaves like the permittivity $\epsilon$, i.e., it couples only to the electric field components that appear in the considered polarization. For instance, in the TM-$k_x$ polarization, the electric field is $\mv E = (E_x,0,E_z)$ and couples to the double $(\epsilon_x,\epsilon_z)$ and to $(\gamma_x,\gamma_z)$, while the magnetic field $\mv H = H_y \hat{\mv{e}}_y$ couples to $\mu_y$.
\begin{center}
\captionof{table}{Relevant material parameters that couple to light depending on the polarization and the incidence plane.} \label{table:materialparamters}
    \begin{tabular}{|c|c||c|c|}
    \hline
         \multicolumn{2}{|c||}{$P=0$ (TE)}&\multicolumn{2}{c|}{$P=1$ (TM)}\\
         \hline\hline
         $(k_x,0,k_z)$ & $(0,k_y,k_z)$ & $ (k_x,0,k_z)$ & $(0,k_y,k_z)$ \\
         \hline
          $\epsilon_y$ &  $\epsilon_x$  & $\epsilon_x$  &  $\epsilon_y$ \\
          $\mu_x$ & $\mu_y$  & $\epsilon_z$  &  $\epsilon_z$ \\
          $\mu_z$ &  $\mu_z$ & $\mu_y$  &  $\mu_x$ \\
          $\gamma_y$ & $\gamma_x$ & $\gamma_x$  &  $\gamma_y$ \\
             -   &  -  &  $\gamma_z$  & $\gamma_z$ \\
         \hline
    \end{tabular}
\end{center}
Here, we shall restrict our attention to the TM-$k_x$ illumination, since in this polarization the fishnet metamaterial that we investigate in Sec.~\ref{subsec:fishnet} shows interesting resonances leading to a negative index. Furthermore, the basic structure made from dielectric spheres on a cubic lattice has a clear spectral signature in terms of the Brewster effect in this polarization. We show in Sec.~\ref{subsec:spheres} that the Brewster angle is captured more accurately using the nonlocal approach. Considering the complexity and the nonlinearity of Eq.~\eqref{eq:dispSSDTM} and of Eqs.~\eqref{eq:reflection-definition} and \eqref{eq:transmission-definition}, the retrieval cannot be performed by inverting and solving ${\rho^{\text{TM}}(k_0,k_x,\epsilon,\mu,\gamma)\overset{!}{=}\rho^{\text{REF}}(k_0,k_x)}$ and ${\tau^{\text{TM}}(k_0,k_x,\epsilon,\mu,\gamma)\overset{!}{=}\tau^{\text{REF}}(k_0,k_x)}$. For this reason, the retrieval is based on fitting the analytically derived reflection and transmission formulas, i.e., Eqs.~\eqref{eq:reflection-definition} and \eqref{eq:transmission-definition} to the referential reflection and transmission coefficients ($\rho^{\text{REF}}$ and $\tau^{\text{REF}}$) of the reference material. The latter quantities can either be measured in experiments or obtained numerically, for example by using the Korringa-Kohn-Rostoker method (KKR) or the Fourier Modal Method (FMM). In this fitting procedure we consider each frequency individually, as the material parameters are functions of the frequency. 
\begin{figure*}[t]
	\centering
  \includegraphics[width=0.81\textwidth]{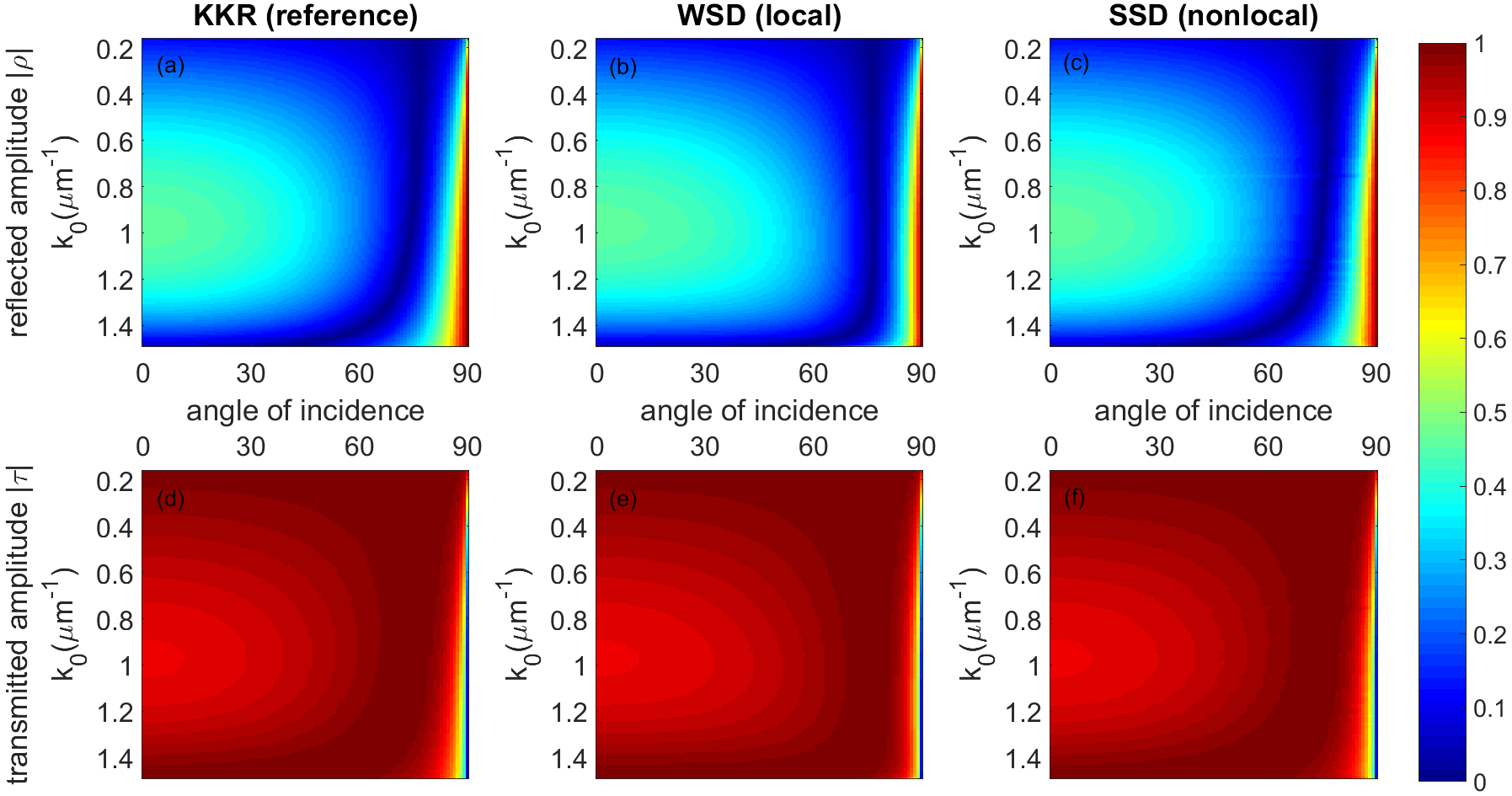}  
	\caption{(a)-(c) amplitude of the reflected light $|\rho|$ and (d)-(f) of the transmitted light $|\tau|$ from a layer of dielectric spheres with thickness $\dslab =1\mu\mathrm{m}$  using different approaches. The left figures ((a) and (d)) correspond to the full-wave simulation of the actual slab as done with the KKR. This can be considered as the reference data. The centered figures ((b) and (e)) are the fitted reflection and transmission amplitudes from a homogeneous slab with the same thickness as the heterogeneous slab using the WSD, i.e., the local approach. The figures on the right ((c) and (f)) are obtained from considering a homogeneous slab with SSD, i.e., retaining nonlocal effects in the effective description. The figure indicates the improvement in capturing the reflection and transmission of the reference material ((a) and (d)) using SSD (nonlocal) compared to WSD (local).}
		\label{fig:randt_full_spheres}
\end{figure*} 
Given reflection and transmission as a function of the angle of incidence, one can minimize a merit function $\delta$ w.r.t. the effective material parameters to capture the reflection and transmission coefficients of the structure. The merit-function $\delta$ is a measure of how well a model applies to homogenize a structure. It rewards the ability of the constitutive relation to recover the electromagnetic response of the slab and is explicitly defined as follows:
\begin{align}
\delta(k_0) = \min_{\epsilon,\mu,\gamma} \sum_{k_x=0}^{k_0}& \frac{w(k_x)}{2} \Bigg( \left| 1- \frac{\rho^{\text{TM}}(k_0,k_x,\epsilon,\mu,\gamma)}{\rho^{\text{REF}}(k_0,k_x)} \right| \nonumber \\ &+  \left| 1- \frac{\tau^{\text{TM}}(k_0,k_x,\epsilon,\mu,\gamma)}{\tau^{\text{REF}}(k_0,k_x)} \right| \Bigg)   \,, \label{eq:meritfunctionrandt}
\end{align}
where $w(k_x)$ is a weight function that is centered in the paraxial regime, i.e., $k_x\approx 0$. We chose here an exponentially decaying dependency, such that
\begin{equation}
    w(k_x) = \ee^{-\alpha k_x}\,,
\end{equation}
with $\alpha = 2.5 \Lambda_x$, with $\Lambda_x$ the lateral period of the structure. The notation $\sum_{k_x=0}^{k_0}$ means the sum over some reasonable finite range of values of $k_x$ belonging to $[0,k_0)$. Below, this range of values will be specified. The purpose of giving more importance to the weight function in the paraxial regime is our requirement to be able to reproduce at least reflection and transmission at normal incidence and look afterwards how far we can stretch this regime of applicability also to oblique incidence angles. As the fit is done at each frequency individually, we add a penalty term to the merit function $\delta$ if there is any spectral discontinuity in any of the effective material parameters. This improves spectral continuity of the effective material parameters. This step renders the final results physical, since the material parameters have to be analytic functions of frequency. 

Using this approach, we retrieve by minimizing the merit function at each frequency individually the material parameters that can describe best the optical response from the original structure as a function of the angle of incidence. The retrieval has been done individually for both WSD and SSD. In our simulations we use $240$ frequencies $k_0$ ranging from $k_{0,\mathrm{min}}$ to $k_{0,\mathrm{max}}$. For every frequency, we simulate $100$ angles of incidences defined as $k_x=k_0\sin(\theta)$, where $\theta$ is the angle of incidence, ranging from $0^{\circ}$ to $89.99^{\circ}$. $90^{\circ}$ is excluded due to numerical reasons. 

\subsection{Basic example: Dielectric spheres on a cubic lattice}\label{subsec:spheres}
Prior the investigation of the complicated, anisotropic fishnet metamaterial, we would like to discuss an easier example of a metamaterial made by isotropic unit cells. For instance, let's consider a metamaterial made from dielectric spheres on a cubic lattice surrounded by vacuum. The spheres shall have a permittivity of $\epsilon{_\mathrm{SPH}}=16$, e.g., Germanium spheres. The assumption of a non-dispersive (constant) permittivity leads to the scalability of Maxwell's equation. Therefore, the critical and relevant length scales are not their magnitudes, but rather their length relative to the spatial periodicity of the unit cell. Without loss of generality, we assume the period $a:=\Lambda_x=\Lambda_y=\dslab=1~\mu\mathrm{m}$. The radius of the spheres shall be $0.45a$ and the free-space wavelengths of the incident light $\lambda \in (4a,40a)\mu\mathrm{m}$. The factor $4$ ensures that the wavelengths in the medium remains longer than the period. Hence a sub-wavelength scenario holds and homogenization is feasible. 

Considering that we are dealing with spherical objects, it is appropriate to use the Korringa-Kohn-Rostoker method (KKR)\cite{stefanou1998heterostructures} to predict reflection and transmission from a slab. The scattered fields from such structures are expressed as a superposition of plane waves and numerically the reflection and transmission amplitude coefficients are solved for. These are depicted in Fig.~\ref{fig:randt_full_spheres}(a) and (d), respectively as a function of the wavenumber $k_0$ and the angle of incidence. We consider here a single layer of the metamaterial.

\begin{figure}[h!]
	\centering
  \includegraphics[width=0.35 \textwidth]{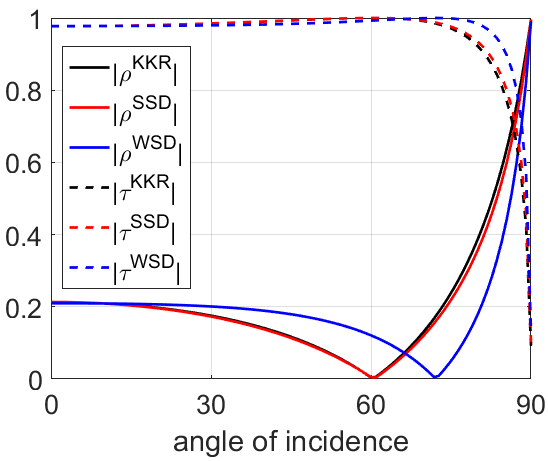}  
	\caption{Reflection and transmission coefficients at a selected frequency of $k_0 = 0.9\frac{\pi}{2a}$ of a metamaterial made from dielectric spheres on a cubic lattice. Geometrical and material details are indicated in the main body of the text. The solid (dashed) curves represent the real (imaginary) part. The reference curves obtained from the KKR are in black. The blue curves are obtained from considering WSD. The red ones for the case of SSD. The red curves are obtained from fitting Eqs.~\eqref{eq:reflection-definition} and \eqref{eq:transmission-definition} to the reference curve. They show a good agreement up to $90^\circ$. Meanwhile, the blue curves, which are obtained from WSD, are showing only an agreement within the paraxial regime.}
	\label{fig:RandT_TMxkx_wide_spheres}
\end{figure} 
 We notice that the dark line in Fig.~\ref{fig:randt_full_spheres}(a) refers to angles with a vanishing reflection, i.e., the Brewster angle as usually encountered for TM polarisation. Due to the isotropy and the lossless character of the constituents, the fitting procedure is restricted to a 2-dimensional or 3-dimensional optimization problem, where the free parameters are the real numbers $\epsilon$, $\mu$  in the case of WSD and additionally for the SSD we consider the nonlocal parameter $\gamma$. Later, for the fishnet metamaterial, we have to consider a 6-dimensional and 10-dimensional optimization scenario for the local and the nonlocal approaches, respectively.

The fitted reflection and transmission coefficients using WSD(Fig.~\ref{fig:RandT_TMxkx_wide_spheres}(b) and (e)) and SSD(Fig.~\ref{fig:randt_full_spheres}(c) and (f)) are depicted as well. Given the simplicity of the structure, it is not surprising that we manage to properly capture the reflection and transmission with both WSD and SSD for small angles of incidence. However, with increasing angles of incidence and increasing wavenumber $k_0$, we observe that only the SSD remains consistent with the reference, while the WSD fails to retrieve the Brewster angle (dark line in the reflection coefficient). This aspect is elucidated more deeply in Fig.~\ref{fig:RandT_TMxkx_wide_spheres}, where we show reflection and transmission coefficients at the selected frequency of $k_0= 0.9 \frac{\pi}{2a}$. We observe that for small angles of incidence, both WSD and SSD match with the reference. Leaving the paraxial regime, we see that only the SSD model captures the reflection coefficient and most importantly, the Brewster angle, whereas the WSD model fails.

\begin{figure}[t]
 \includegraphics[width=0.47\textwidth]{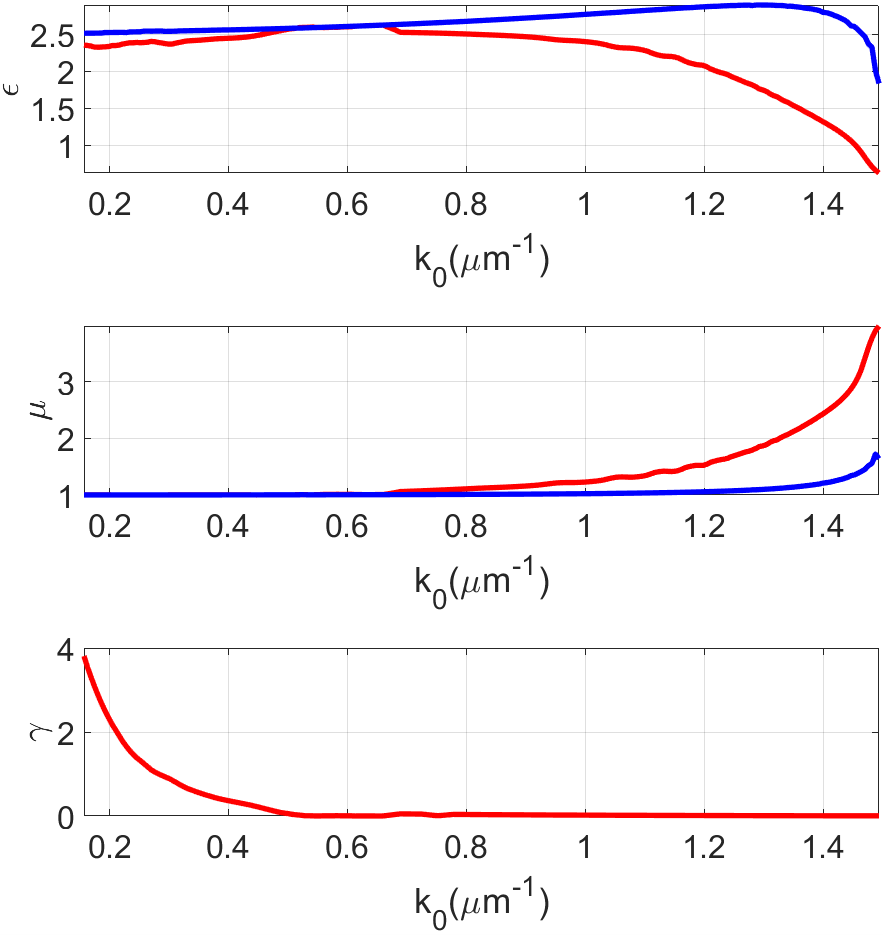}
	\caption{(Colors online) Effective permittivity $\epsilon$, effective permeability $\mu$, and the effective nonlocal parameter $\gamma$ as a function of the frequency $k_0$ using both local (blue) and nonlocal (red) approaches for the metamaterial made from a layer of dielectric spheres arrange on a cubic lattice. The results are obtained from fitting reflection and transmission coefficients \eqref{eq:reflection-definition} and \eqref{eq:transmission-definition} by means of absolute deviations from the exact data as defined in Eq.~\eqref{eq:meritfunctionrandt}. }
	\label{fig:effectivematerialparamters_spheres}
\end{figure}

The retrieved effective material parameters are shown in Fig.~\ref{fig:effectivematerialparamters_spheres}. For very small frequencies, we obtain an effective permittivity close to the value predicted form the Maxwell-Garnett formula for the quasi-static approximation
\begin{equation}
\epsilon_{\mathrm{MG}} = \epsilon_{\mathrm{b}}+3f\epsilon_{\mathrm{b}}\frac{\epsilon_{\mathrm{SPH}}-\epsilon_{\mathrm{b}}}{\epsilon_{\mathrm{SPH}}+2\epsilon_{\mathrm{b}}-f(\epsilon_{\mathrm{SPH}}-\epsilon_{\mathrm{b}})} \approx 2.40\,.
\end{equation}
where $f= \frac{4\pi}{3}0.45^3\approx 0.38$ is the filling factor of the inclusion with $\epsilon_{\mathrm{SPH}}=16$ in the background with $\epsilon_{\mathrm{b}}=1$. Additionally, in the small frequency regime we observe the rather trivial magnetic permeability $\mu=1$, as expected. Only at higher frequencies, both models show different effective material parameters and a dispersive permeability $\mu$ occurs. We note that the nonlocal material parameter has a dependency that is proportional to $\propto k_0^{-4}$. This is not surprising, considering its emergence in the fourth order of the Taylor expansion. We do not observe such proportionality for the permeability, as the term emerging from the second order term in the Taylor expansion ($\alpha$ in Eq.~\ref{eq:const-relations}) had been already suitably scaled. However, except this rather trivial scaling the term is rather small but non-negligible. It has to be considered to explain properly the optical response from the considered metamaterial; even at small frequencies.

\newpage
\subsection{Fishnet metamaterial}\label{subsec:fishnet}

\begin{figure}[t]
	\centering
  \includegraphics[width=0.33\textwidth]{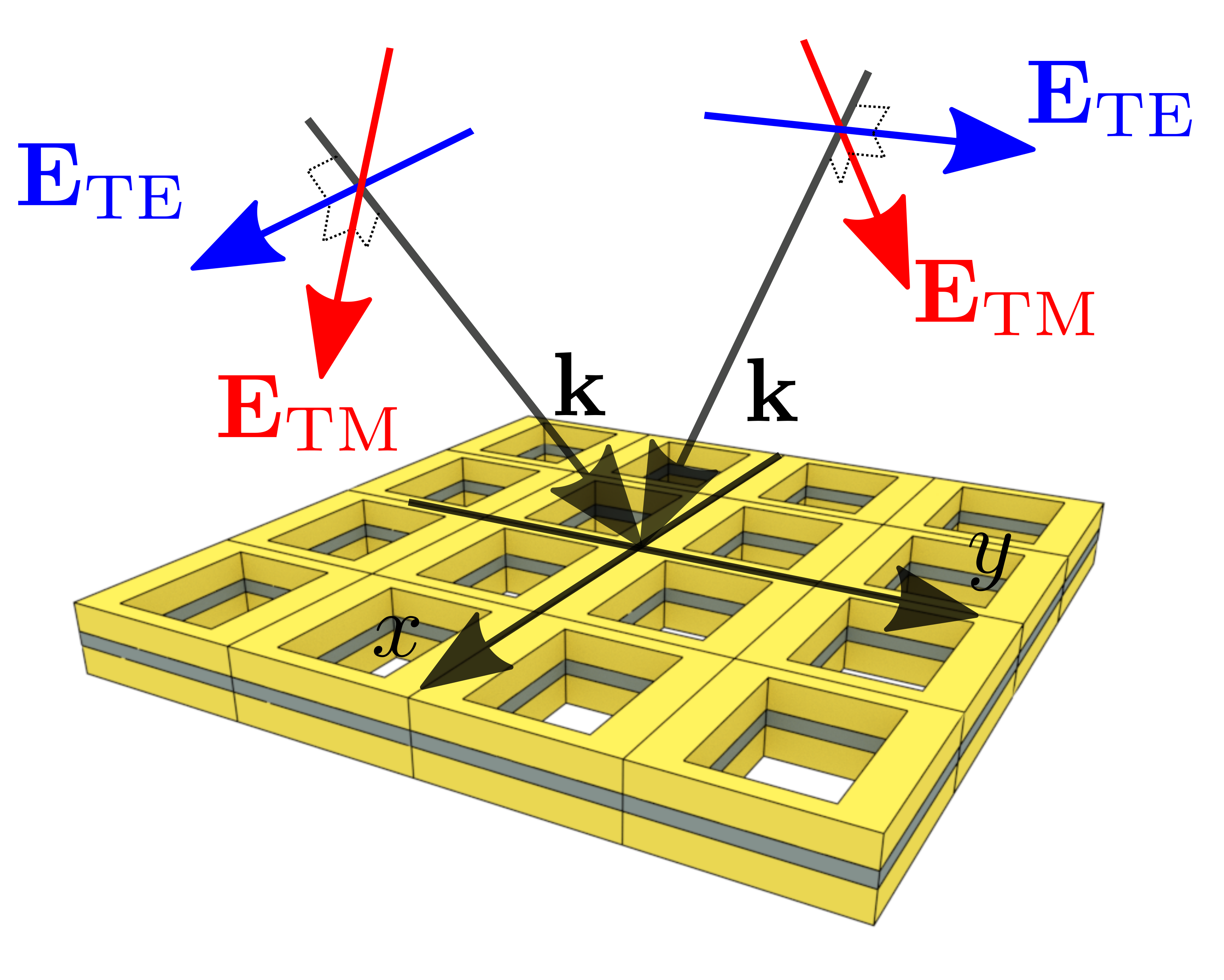}
	\caption{Fishnet metamaterial. We consider a biperiodic structure with periods $\Lambda_x=\Lambda_y=600\,$nm and consider an extension of the thin film in $z$ direction of $200\,$nm. The fishnet consists of rectangular holes with the width $w_x= 100\,$nm and $w_y= 316\,$nm. They consist in a stack of layers made of two $45\,$nm Ag layers separated by a thin dielectric spacer, $30\,$nm of MgF$_2$, with $n_{\text{MgF}_2}=1.38$. The rest is filled with air\cite{PhysRevB.81.035320}.}
	\label{fig:fishnet}
\end{figure}

As an exemplary metamaterial of current interest, which we consider in the following retrieval, we choose the fishnet metamaterial. It exhibits a negative refraction in a specific frequency range with both a dispersive permittivity and a dispersive permeability. The basic geometry is shown in Fig.~\ref{fig:fishnet}. The geometrical parameters are taken from literature \cite{FISHNETGEOMETRY}. It consists of a centro-symmetric unit cell with side lengths of $\Lambda_x=\Lambda_y=600$ nm being replicated in the $xy$-plane and a stacking of air-metal-dielectric-metal-air nanowires in $z$-direction with a total thickness of $200$ nm. The nanowires are perpendicularly aligned and form rectangular holes with widths of $w_y=100$ nm and $w_x=316$ nm. The metal (silver) layers have a thickness of $45$ nm. Silver is described by a Drude model for the permittivity that reads 
\begin{align*}
    \epsilon_{\text{Ag}}= 1 - \frac{\omega_p^2}{\omega^2+\ii \Gamma \omega}\,,
\end{align*}
with the plasma frequency $\omega_p=13700$ THz and the relaxation rate $\Gamma=85$ THz. The silver layers are separated by a nondispersive magnesium fluoride spacer with $\epsilon_{\text{MgF}_2}=1.9044$ and a thickness of $30$ nm. This metamaterial admits a negative index in the TM-$k_x$ polarization for frequencies around $k_0 = 4.3~\mu\text{m}^{-1}$. Since we are interested in this negative index property, we perform the retrieval for this polarization. The retrieval procedure for the other three illumination directions advances similarly and will not be shown here, as we are mainly interested in the material parameters that are linked to the negative index behaviour. 

\begin{figure*}[t]
	\centering
  \includegraphics[width=0.81\textwidth]{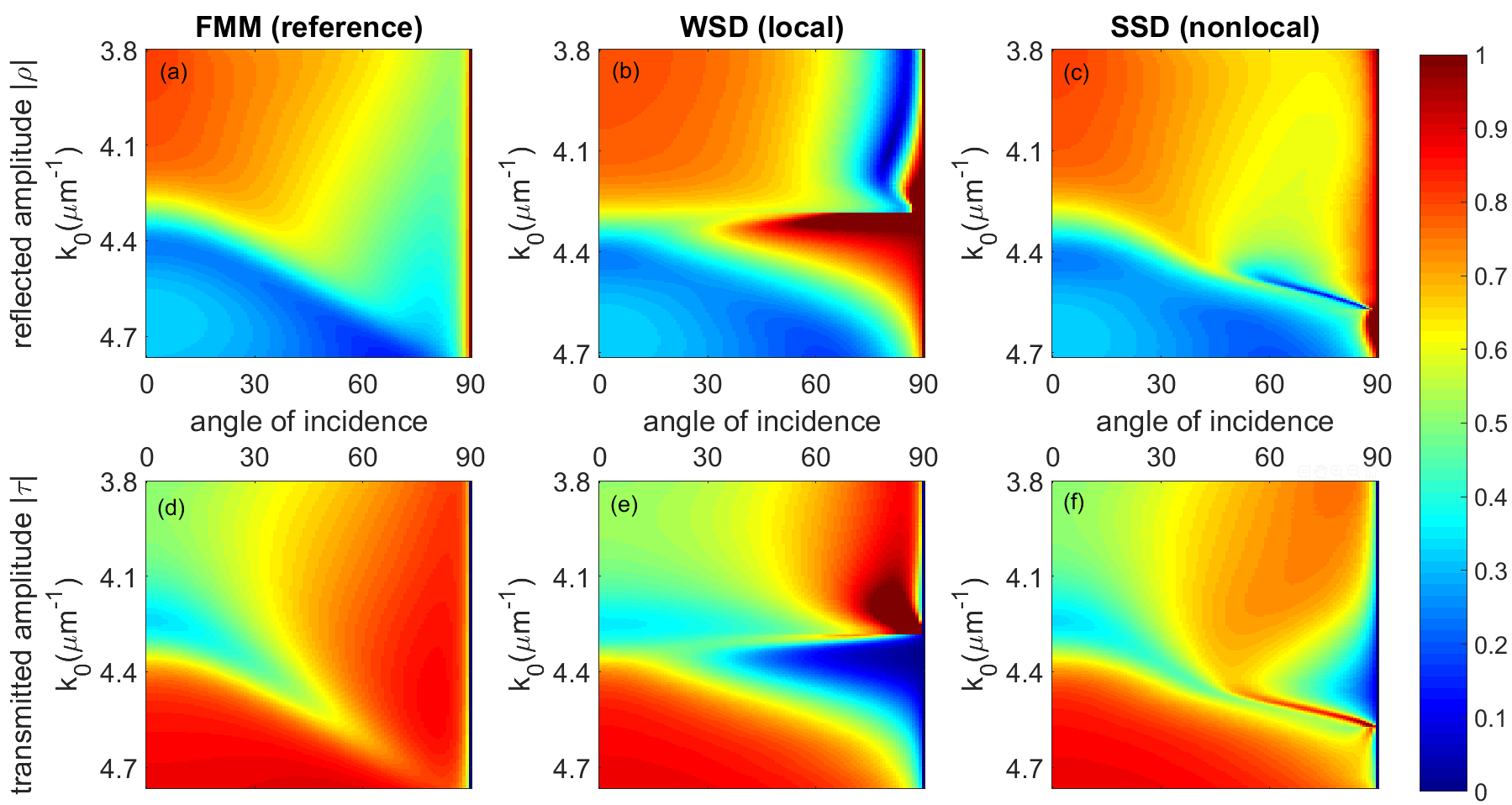}  
	\caption{(a)-(c) amplitude of the reflected light $|\rho|$ and (d)-(f) of the transmitted light $|\tau|$ from one fishnet layer with thickness $\dslab =200$ nm using different approaches. The left figures ((a) and (d)) correspond to the full-wave simulation of the actual fishnet slab as done with the FMM. This can be considered as the reference data. The centered figures ((b) and (e)) are the fitted reflection and transmission amplitudes from a homogeneous slab with the same thickness as the fishnet using the WSD, i.e., the local approach. The figures on the right ((c) and (f)) are obtained from considering a homogeneous slab with SSD, i.e., retaining nonlocal effects in the effective description. The figure indicates the improvement in capturing the reflection and transmission of the reference material ((a) and (d)) using SSD (nonlocal) compared to WSD (local).}
		\label{fig:randt_full}
\end{figure*}

To obtain the numerical data for reflection and transmission from a slab of the fishnet, we perform a full-wave simulation by using a Fourier Modal Method (FMM). In this method, we expand the eigenmodes of the $xy$-periodic structure into Bloch modes and in $z$-direction into plane waves for wavelengths $\lambda$ in the near IR-range with $\lambda \in (\frac{2\pi}{4.768},\frac{2\pi}{3.8})~\mu\text{m}$. Outside the metamaterial the fields are expanded into plane waves. By matching the interface conditions between substrate and cladding and the individual layers that form the fishnet metamaterial, respectively, all amplitudes of all modes are found. In the numerics, a sufficient large number of modes has been taken into account to achieve convergent results for reflection and transmission.     
Since the structure is sub-wavelength, only the $0^{\text{th}}$-diffraction order in transmission and reflection is propagating and will carry energy into the cladding ($z>\dslab$) or the substrate ($z<0$), respectively. Higher-order diffraction contributions are, therefore, suppressed and homogenization is feasible. The $0^{\text{th}}$-order reflection and transmission coefficients obtained from a fishnet slab with $\dslab=200~\text{nm}$ will be represented by $\rho^{\text{FMM}}$ and $\tau^{\text{FMM}}$, respectively and are shown in Fig.~\ref{fig:randt_full} (a) and (d), for $k_0 \in (3.8,4.768)~\mu\text{m}^{-1}$ and the angle of incidence from $0^\circ$ to $90^\circ$, meaning $\forall k_0 \in (3.8,4.768)~\mu\text{m}^{-1}: k_x \in (0, k_0)$. Only the amplitudes are shown but of course the values are complex. These are the amplitudes that have to be reproduced in the homogeneous description.

\begin{figure} [H]
\centering
  \includegraphics[width=0.362 \textwidth]{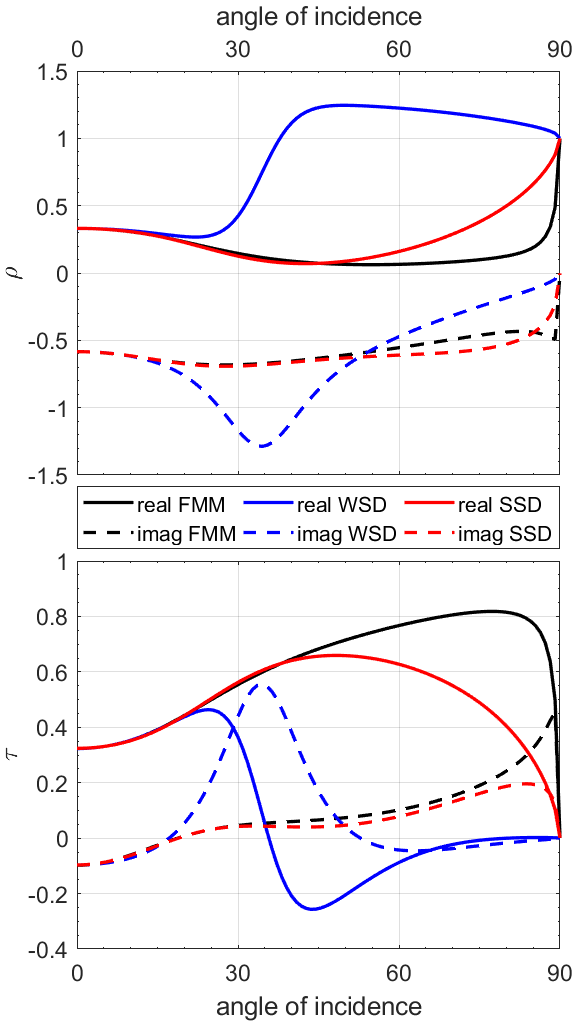}  
	\caption{Reflection (top) and transmission (bottom) coefficients from one fishnet layer at a selected frequency of $k_0 = 4.2414~\mu\text{m}^{-1}$. The solid (dashed) curves represent the real (imaginary) part. The reference curves obtained from the FMM are in black while the blue curves are obtained from considering WSD and the red ones for the case of SSD. The red curves are obtained from fitting Eqs. \eqref{eq:reflection-definition} and \eqref{eq:transmission-definition}
to the reference curve and show a good agreement up to $50^\circ$. Meanwhile, the blue curves, which are obtained from WSD, are showing only an agreement within the paraxial regime.}
	\label{fig:RandT_TMxkx_wide}
\end{figure}

The reflection and transmission coefficients evaluated with the retrieved effective properties are shown in Fig.~\ref{fig:randt_full}. Evidently, it reveals the improvement in capturing the reflection and transmission of the reference material ((a) and (d)) using SSD ((c) and (f)) compared to WSD ((b) and (e)). The reflection and transmission for a selected frequency of $k_0 = 4.2414~\mu\text{m}^{-1}$ are depicted in Fig.~\ref{fig:RandT_TMxkx_wide}.

From these results (Figs.~\ref{fig:randt_full} and particularly  Fig.~\ref{fig:RandT_TMxkx_wide}) we mainly observe two outcomes. With retaining nonlocality we can not only increase the agreement with the reference curves, but also the functional behavior, i.e., the curvature of reflection and transmission w.r.t. the angle of incidence, seems to be more realistic than in the case when WSD only is considered. This is a solid confirmation in favor for the relevance of including nonlocal material parameters into the effective description of metamaterials.

\begin{figure}[H]
  \includegraphics[width=0.49\textwidth]{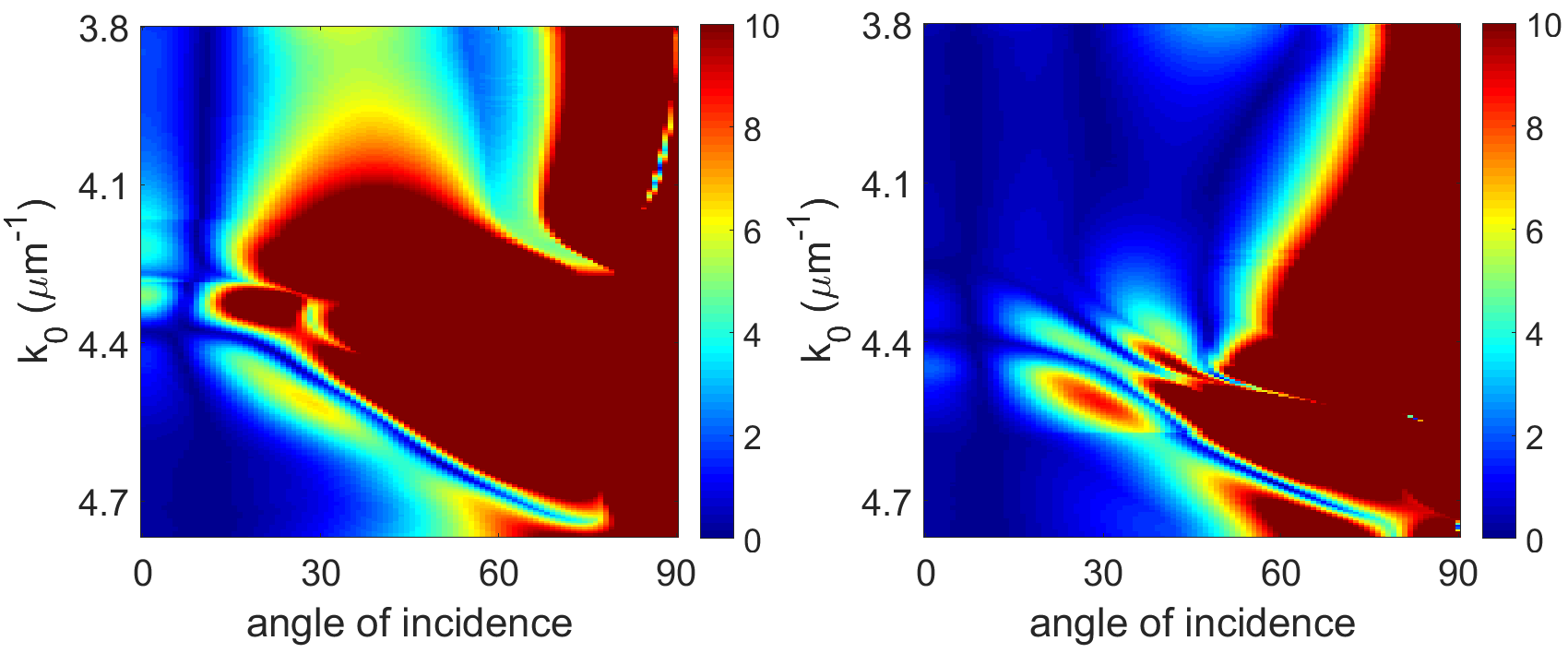}  
	\caption{Deviations from the reflection and transmission coefficients in \% for (left) the local and (right) the nonlocal approaches for frequencies $k_0 \in [3.8,4.768]~\mu\text{m}^{-1}$ and the angle of incidence from $0^\circ$ to $90^\circ$. In comparison to the local approach, the nonlocal one covers a larger parameter space (blue region) where the homogenization is meaningful and effective material parameters can be retrieved. The colorbar is truncated to $10\% $ to indicate a threshold of applicability. }
	\label{fig:deviationsfromrandt}
\end{figure}

To quantify the findings and to allow for a better discussion, we show in Fig.~\ref{fig:deviationsfromrandt} in \% the absolute deviation between the amplitude of reflection and transmission as calculated with the FMM to those fitted using either the WSD (Fig.~\ref{fig:deviationsfromrandt} (left)) or the SSD model (Fig.~\ref{fig:deviationsfromrandt} (right)), respectively. For a better discussion of the deviations, the color axis has been truncated to 10\%. The figures show two regimes of interest. The blue regime is the regime where reflection and transmission can be captured quite well with the retrieved material parameters and, hence, the homogenization is meaningful. In contrast, the red regions refer to deviations above $10\%$ from the reference, i.e., the region where the model fails to capture the electromagnetic response adequately which leads to the failure of the homogenization approach. We clearly see that for all frequencies $k_0$, the nonlocal approach pushes the agreement to higher incidence angles. Of course, as we also expected, around the resonance frequency $k_0 = 4.3~\mu\text{m}^{-1}$ the homogenization becomes less reasonable. Accordingly, in the blue regions of Fig.~\ref{fig:deviationsfromrandt} the metamaterial can be effectively characterized.

The effective material parameters that were obtained in the retrieval are shown in Fig.~\ref{fig:effectivematerialparamters} as solid lines. The parameters that appear in both the local and the nonlocal constitutive relations that result from the retrieval with the WSD and the SSD are shown within one figure. The parameters that appear only in the nonlocal constitutive relation are shown alone. There are a few things worth to discuss.    

First of all, using the local approach, we note that the permittivity $\epsilon_{x}$ has an anti-Lorenzian shape around the resonance frequency $k_0 = 4.3~\mu\text{m}^{-1}$, leading to a negative imaginary part. The permittivity $\epsilon_{x}$ therefore has a complex pole in the upper complex $k_0$ half-plane, and hence, violates causality \cite{PhysRevB.84.054305}. This unphysical anti-resonance is no longer existing when the nonlocal constitutive approach is considered. Moreover, as physically expected, the permittivity $\epsilon_{x}$ shows a Drude type behaviour. This is the usual response expected for a diluted metal; which the fishnet metamaterial is actually. Second, the permeability $\mu_{y}$ is in both models nearly identical and shows a Lorentzian functional dependency. This behavior is expected due to the structure of the fishnet. When light in TM-$k_x$ polarization couples to the metallic nanowires, circular currents are induced and a magnetization occurs. The magnetization is driven into resonance at $k_0 = 4.3~\mu\text{m}^{-1}$ and the Lorentzian dispersion is centered around that frequency. 
Third, the parameter that is hard to interpret is the $z$-component of the permittivity, $\epsilon_{z}$. It differs in both models WSD and SSD. This parameter exhibits a quite strange behavior at some frequencies where $\Im( \epsilon_{z})$ is negative in both the WSD and the SSD. However, we stress that this parameter is hard to capture in the present geometry. At normal incidence, for example, the electromagnetic field does not couple at all to this $z$-component. If only the response at normal incidence is considered, this component cannot be retrieved. At oblique incidence the situation improves somewhat, but after all it turns out to be rather insensitive. This can be explained by the fact that the wavenumber in the metamaterial is quite larger. Hence, even though excited at oblique incidence from the surrounding, the plane waves that are the eigenmodes in the metamaterial propagate inside in a paraxial direction. Therefore, they do not fully probe the $z$-component of the permittivity tensor. This will be discussed later and explained in Fig.~\ref{fig:sensitivity}.
Fourth, considering the nonlocal parameters $\gamma_{x}$ and $\gamma_{z}$, they show a resonance at a frequency around ${k_0=4.3~\mu\mathrm{m}^{-1}}$, where the negative index has it's minimum. Note that the nonlocal parameters are always at least one order of magnitude smaller than the local parameters. Ways how to design metamaterials such that $\gamma$ is maximized is still an open question and subject to study. Nevertheless, we can summarize that with those nonlocal material parameters we can significantly improve our ability to describe the optical response from the considered fishnet metamaterial in a slab geometry, as mainly evidenced in Fig.~\ref{fig:deviationsfromrandt}. 
\subsection{Sensitivity and robustness analysis}\label{subsec:sensitivity-analysis}
\begin{figure}[H]
\centering
 \includegraphics[width=0.47\textwidth]{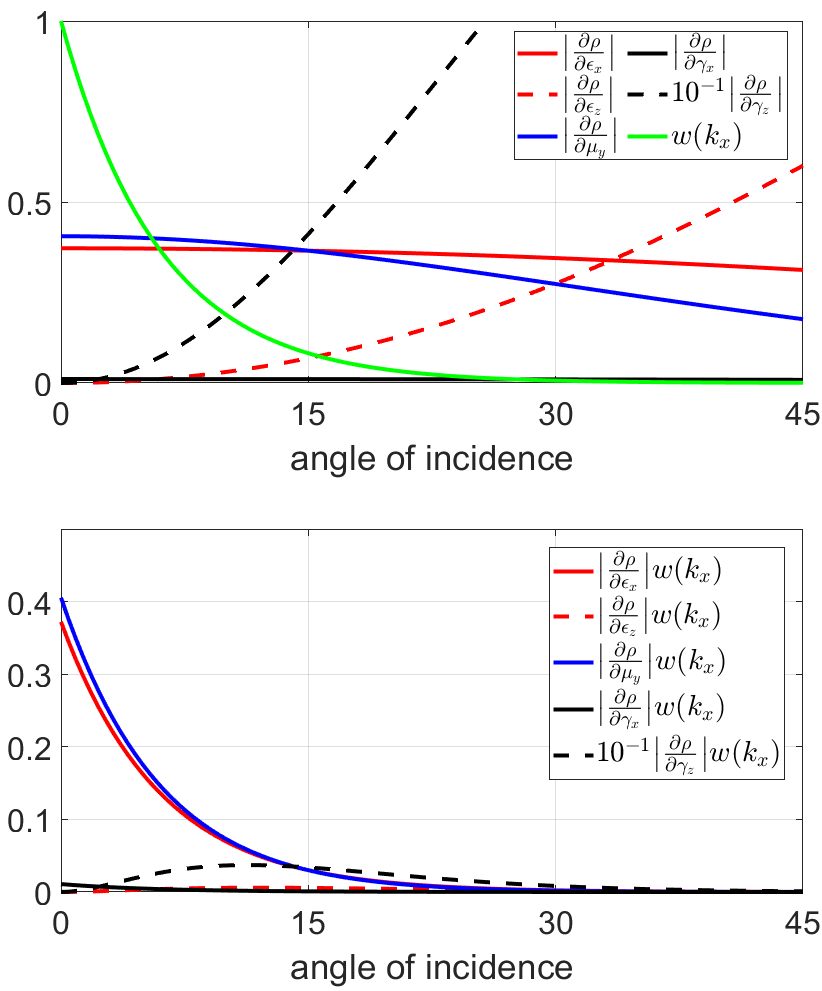}
	\caption{(Color online) Sensitivity of the reflection coefficient $\rho$ w.r.t. effective material parameters and weight function $w(k_x)$ used in the retrieval procedure. The partial derivatives are evaluated at the resonance frequency of the fishnet. The upper figure shows the absolute values of the derivative and the weight function. The bottom figure reveals the effective sensitivity in the retrieval procedure, where the partial derivatives are weighted with $w(k_x)$.}
	\label{fig:sensitivity}
\end{figure}

To evaluate a bit more in detail why some specific material parameters can be reliably retrieved while others not, we performed a sensitivity analysis. The analysis is presented here for the reflection coefficient but identical conclusions can be obtained when doing the analysis with the transmission coefficient. The sensitivity of the reflection coefficient on the material parameters is shown in Fig.~\ref{fig:sensitivity}. We investigate for this purpose the Jacobi-matrix of the reflection coefficient w.r.t. the material parameters, i.e., partial derivatives 
\begin{align}
J_{\rho}= \left(\frac{\partial\rho}{\partial p} \right)\,,
\end{align}
where $p$ is the set of the effective material parameters $p=\{\epsilon_x,\epsilon_z,\mu_y,\gamma_x,\gamma_z\}$. The amplitude of the partial derivatives shall serve as measure of sensitivity. They express how much does the analytical reflection coefficient $\rho$ change by varying one of the parameters.

\begin{figure} 
 \includegraphics[width=0.49\textwidth]{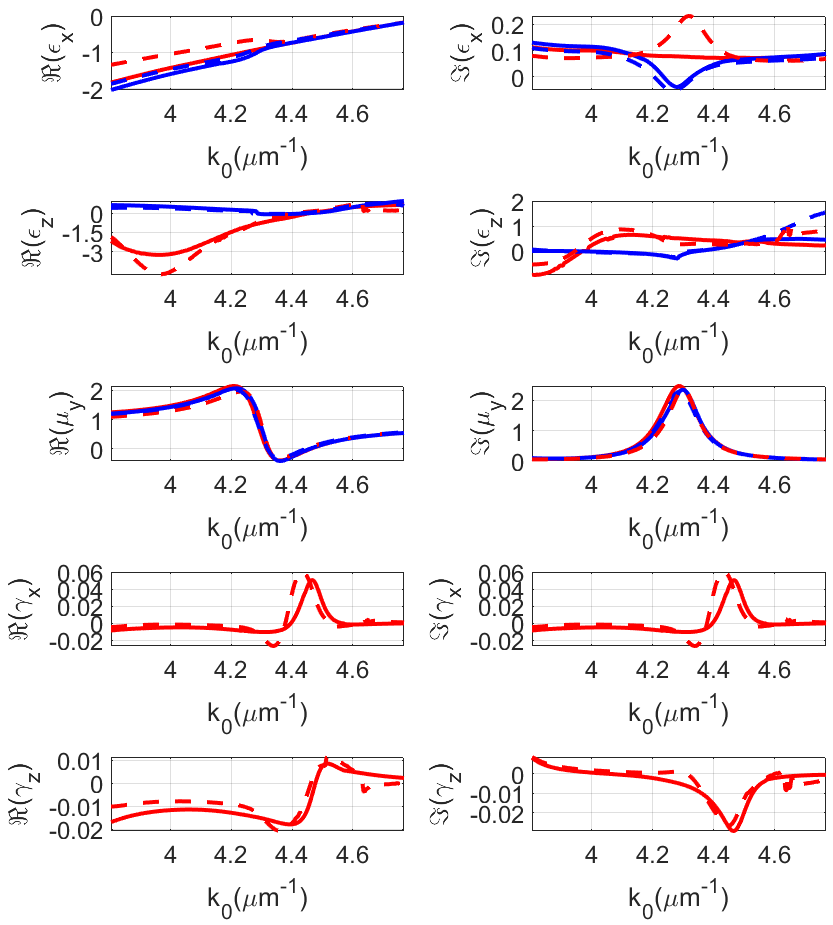}
	\caption{Real and imaginary parts of the effective permittivities $\epsilon_{x}$ and $\epsilon_{z}$, effective permeability $\mu_{y}$, and the effective nonlocal parameters $\gamma_{x}$ and $\gamma_{z}$ as a function of the frequency $k_0$ using both local (blue) and nonlocal (red) approaches. The solid (dashed) lines correspond to the retrieval from one (two) layer(s). The results are obtained from fitting reflection and transmission coefficients \eqref{eq:reflection-definition} and \eqref{eq:transmission-definition} by means of absolute deviations from the exact data as defined in Eq. \eqref{eq:meritfunctionrandt}. }
	\label{fig:effectivematerialparamters}
\end{figure} 

In Fig.~\ref{fig:sensitivity} (top) we show the partial derivatives w.r.t. the real parts of the effective material parameters evaluated at the resonance frequency of the fishnet ${k_0=4.3~\mu\mathrm{m}^{-1}}$. The imaginary parts are not important here, since they only show how much susceptible the reflection coefficient is w.r.t. loss. We are interested into the propagation aspect, i.e., the real parts. We also included the green solid line showing the weighting function we introduced in the fitting procedure. The weighting function has been introduced with the purpose to capture the electromagnetic response at normal incidence at least. In Fig.~\ref{fig:sensitivity} (bottom) we show the effective sensitivity in the fitting procedure. It is the product of $\frac{\partial\rho}{\partial_p}w(k_x)$. Clearly, the fitting is very sensitive w.r.t. $\epsilon_x$ and $\mu_y$, the relevant material parameters at normal incidence, and less sensitive w.r.t. $\epsilon_z$. This explains the difficulty of retrieving this latter parameter. The nonlocal parameters $\gamma_x$ and $\gamma_z$ show different behavior. We can deduce that $\gamma_z$ is a very sensitive parameter and thus important in the retrieval procedure, leading to improvements in retrieving the reflection coefficient of the fishnet. In contrast, $\gamma_x$ seems to be not even important in the paraxial regime. Its sensitivity is very low, i.e., a huge change in the parameter would not affect the reflection and transmission value. Therefore, it is pointless to discuss this parameter.  

\begin{figure} 
  \includegraphics[width=0.34 \textwidth]{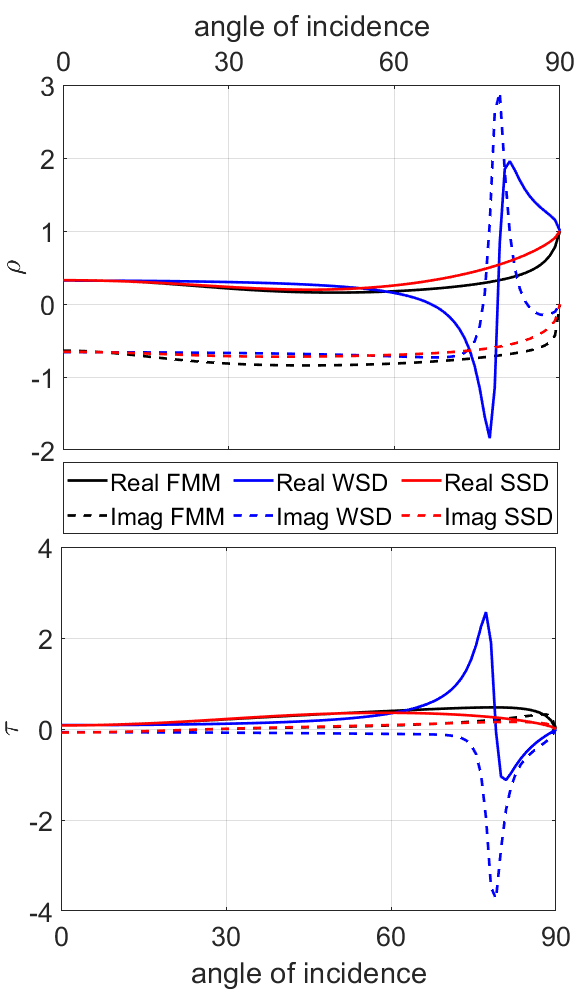}  
	\caption{Reflection (top) and transmission (bottom) coefficients from two fishnet layers at a selected frequency of $k_0 = 4.2414~\mu\text{m}^{-1}$. The solid (dashed) curves represent the real (imaginary) part. The reference curves obtained from the FMM are in black while the blue curves are obtained from considering WSD and the red ones for the case of SSD. The red curves are obtained from fitting Eqs. \eqref{eq:reflection-definition} and \eqref{eq:transmission-definition}
to the reference curve and show a good agreement up to $70^\circ$. Meanwhile, the blue curves, which are obtained from WSD, are showing only an agreement within the paraxial regime.}
	\label{fig:RandT_TMxkx_wide_two_layers}
\end{figure}

Furthermore, to access the liability of the retrieved material parameters, we applied the retrieval procedures to a fishnet metamaterial from two functional layers. We considered the same frequency range and again all angles of incidence. Similarly to the case of a single layer, we show complex reflection and transmission coefficients at a selected frequency of $k_0=4.2414\mu\mathrm{m}^{-1}$ in Fig.~\ref{fig:RandT_TMxkx_wide_two_layers}. It shows that the nonlocal approach (red lines) is more efficient in reproducing the reference data (black lines) than the local approach (blue lines). In this situation, the retrieved effective material parameters are plotted as the dashed lines in Fig.~\ref{fig:effectivematerialparamters}. The effective material parameters seem to be convergent for both WSD and SSD models, rendering our retrieval procedure stable. Only $\epsilon_x$ in the case of SSD (compare red solid and red dashed in Fig.~\ref{fig:effectivematerialparamters}) differs weakly when considered either one or two functional layers. In particular, we encounter a weak Lorentzian behaviour around the magnetic resonance frequency, while in the WSD the unphysical anti-Lorentzian remains. We attribute this additional feature due to coupling effects between consecutive functional layers of the metamaterial.
 
\begin{figure} 
  \includegraphics[width=0.46 \textwidth]{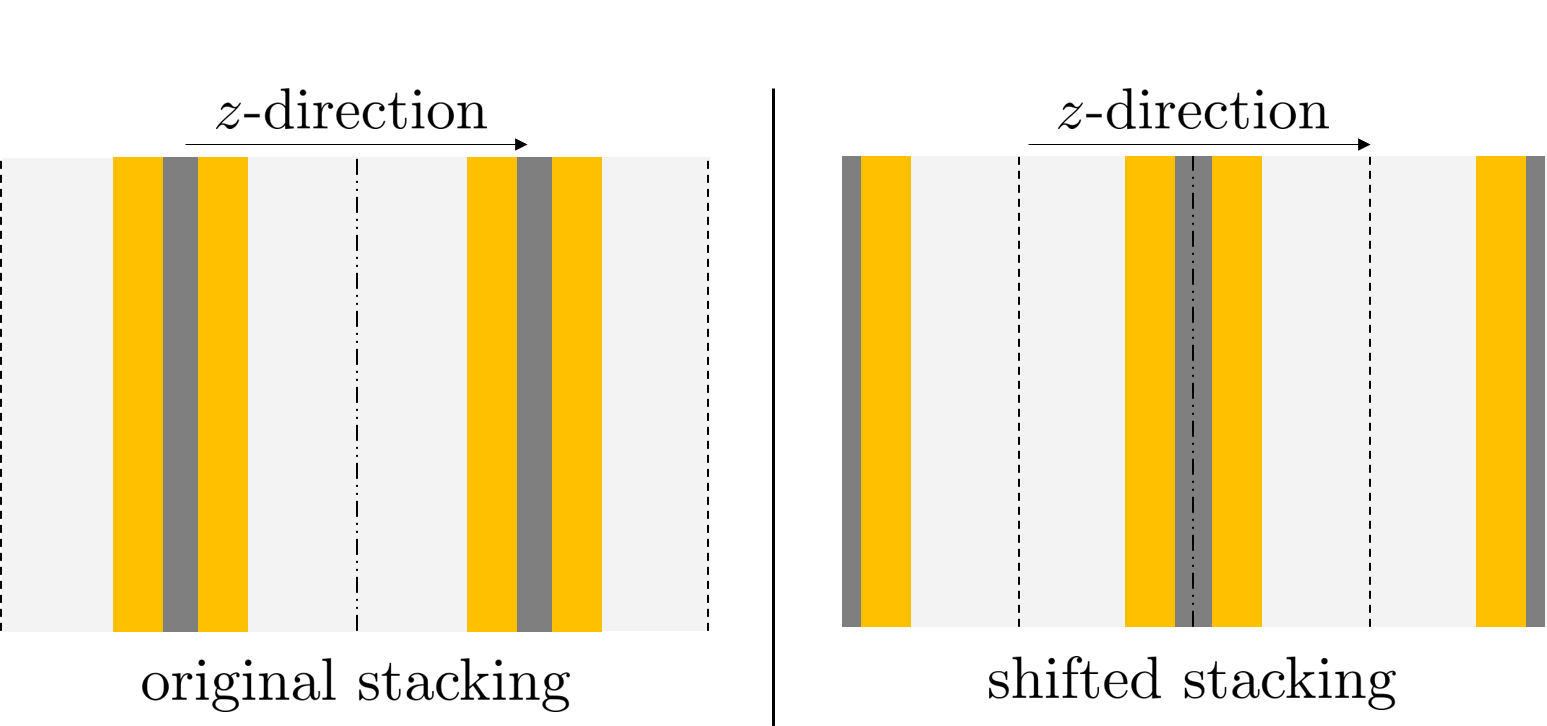}  
	\caption{Side view of two functional fishnet layers in their original (shifted) stacking on the left (right). The unit cells are separated by a dot-dashed line. The dashed lines represent the limit of a single air layer. Yellow signifies the metallic layer, dark grey signifies the dielectric layer, and light grey signifies the vacuum.}
	\label{fig:rearranging_fishnet}
\end{figure}

Moreover, we studied the behavior of the effective material parameters under a redefinition of the unit cell, as illustrated in Fig.~\ref{fig:rearranging_fishnet}. To be specific, the bulk arrangement does not change but now we consider no longer a unit cell with the metal-dielectric-metal structure in the center but rather a unit cell where this material is equally distributed on both edges of the unit cell. So the unit cell is shifted by half a unit cell in the bulk material. It has been pointed out in the past that such redefinition of the unit cell indeed affects the response, i.e. the truncation of the metamaterial is important \cite{PhysRevLett.115.177402}.

\begin{figure} 
 \includegraphics[width=0.49\textwidth]{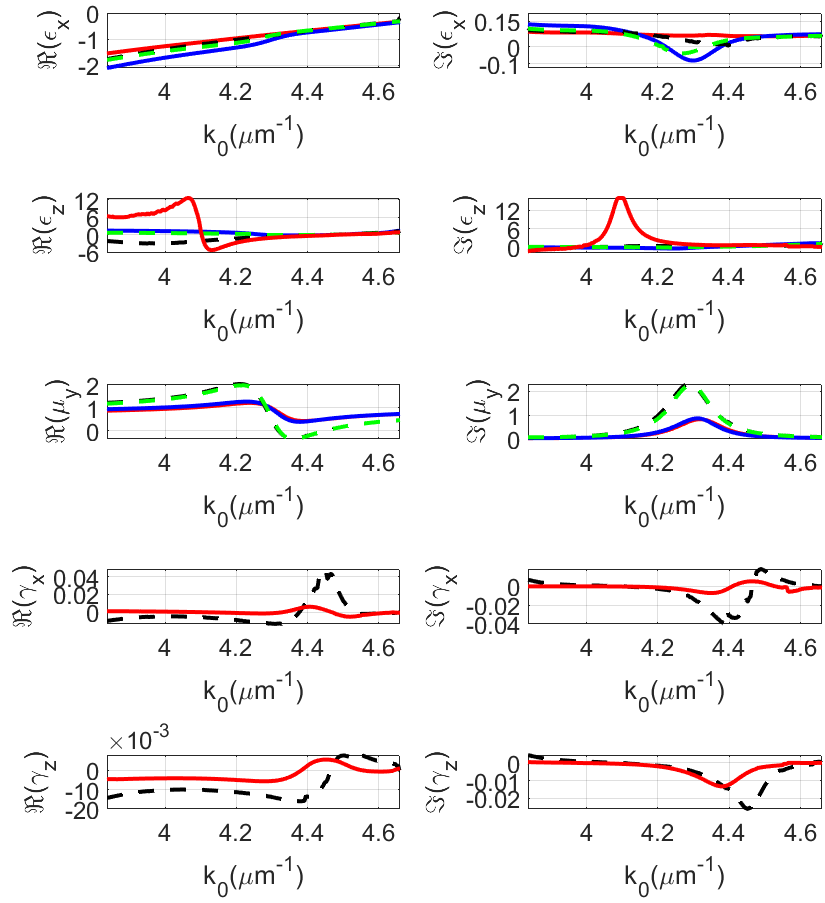}
	\caption{(Colors online) Retrieved real and imaginary parts of the effective material parameters $\epsilon_{x}$, $\epsilon_{z}$, $\mu_{y}$, $\gamma_{x}$, and $\gamma_{z}$ as a function of the frequency $k_0$ for the redefined and the original unit cell as illustrated by solid and dashed lines, respectively. The unit cells correspond to those discussed in Fig.~\ref{fig:rearranging_fishnet}. We considered a slab of the metamaterial made from two unit cells. Red and black (blue and green) curves correspond to the material parameters obtained from the nonlocal (local) approach.}
	\label{fig:effectivematerialparamters_rearranged}
\end{figure}

Similar to the case of the original stacking, we numerically calculate the reflection and transmission coefficient for the shifted stacking. We apply afterwards our procedure to retrieve the effective material parameters. We considered once a metamaterial slab made from a single functional layer but also a metamaterial slab made from two functional layers. The effective material parameters were retrieved and the results are shown in Fig.~\ref{fig:effectivematerialparamters_rearranged} when the slab had been made from two functional layers.

The retrieved results for one layer are largely different when compared to the previously presented effective properties. This is somewhat expected, since for a single functional layer in the alternative choice of the unit cell the induced dipole moments in the two metal layers are spatially separated by a rather thick air layer now. This leads to the vanishing of the effective magnetic response in the frequency range of interest. Their anti-asymmetric mode will be supported at largely different frequencies. The metamaterial behaves effectively only as a diluted metal if the slab is made from a single unit cell only. Hence, no magnetic resonance, i.e., $\mu_y \approx 1$ and a Drude-type permittivity can be seen (results are not explicitly shown here). 

However, the magnetic resonance emerges again as soon as a stack of two functional layers in the $z$-direction is considered for the metamaterial made from the differently defined unit cell. Now, the second metal layer of the first unit cell is only separated by two thin dielectric layers of half size from the first metal layer of the second unit cell. This restores the original geometry at least in the central region of the considered metamaterial slab. This leads to a strong, antisymmetric coupling between the induced dipole moments in the metal layers that causes the Lorentzian-shape of the permeability $\mu_y$. Addiotionally, we would like to emphasize that the amplitude of the magnetic resonance is lower than in the original structure. This anticipated behavior is a consequence of a diluted magnetization inside the two functional layers. Similar behavior holds for the nonlocal parameters $\gamma_x$ and $\gamma_z$. The electric properties seem to be weakly invariant under this shifting. This clearly hints to the fact that the electric response is rather caused by the individual metallic plates and not by their coupling. Only the $\epsilon_z$ parameter changes in the retrieval, where a Lorentz-type resonance emerges at lower frequencies. This might be linked to some specific coupling effects that takes place between the separated metal layers in the unconventional stacking sequence. Anyway, the discussion of this parameter is delicate because it is a weakly sensitive parameter and hard to retrieve in each case, as discussed above.

\section{\label{sec:conclusions}Conclusions}
In this contribution, we investigated the response of two optical metamaterials. First, a basic and isotropic material made from dielectric spheres on a cubic lattice up to frequencies close to the first photonic bandgap. Second, an advanced anisotropic metamaterial, the fishnet metamaterial in the frequency range where it undergoes a resonant coupling with light yielding a negative index behaviour in the TM-$k_x$ polarization. We retrieved both local and nonlocal material parameters emerging in the TM-$k_x$ polarization and sketched the procedure for the other three cases as well in Table \ref{table:materialparamters}. The retrieval was performed by fitting the analytically derived reflection and transmission coefficients for both weak and strong spatial dispersion and compare these to the complex-valued reflection and transmission coefficients of the heterogeneous slab. We clearly see that for all simulated frequencies, the nonlocal approach pushes the agreement to higher angles of incidence than the local approximation and that the reflection and transmission from a slab can be captured more efficiently using the nonlocal approach. In addition, the permittivity $\epsilon_{x}$ shows unphysical behaviour around the resonance frequencies in the retrieval using WSD model. This has been lifted by introducing the nonlocal material parameters. This is another indication that it is important to retain nonlocality for a more realistic homogenization of optical metamaterials. These findings have been obtained using the fishnet structure as a test subject, which sustains a negative index and therefore of utmost importance in applications. However, the procedure can be readily applied to other metamaterials. Moreover, the Fresnel equations above can be also applied for a different incident medium. However, it turned out to be complicated to retrieve consistent material parameters. We understand these difficulties as a consequence of a weak but non-negligible overlap of the localized fields in the vicinity of the interface with the surrounding. If the ambient material changes, these resonances are easily detuned in their spectral positions. For instance, in the case of the Fishnet metamaterial, the plasmonic resonances on which we rely are spectrally detuned if the surrounding material is modified and that shifts the typical frequencies of the different material parameters, i.e., the resonance frequency of the Lorentzian in the permeability or the plasma frequency in the permittivity.

\section*{Acknowledgments}
We gratefully acknowledge financial support by the Deutsche Forschungsgemeinschaft (DFG) through CRC 1173. K.M. also acknowledges support from the Karlsruhe School of Optics and Photonics (KSOP). A.K. is supported by the Austrian Science Fund (FWF) under Project No. M 2310-N32. The authors would like to thank Andreas Vetter for providing reference data for Sec.~\ref{subsec:spheres}.
\bibliography{retrievalbib}

\end{document}